\documentclass[aps,pra,10pt,notitlepage,twocolumn]{revtex4-2}
\usepackage{amsmath}
\usepackage{amssymb}

\usepackage{tikz}
\usetikzlibrary{quantikz}

\usetikzlibrary{calc}

\newcommand{\ignore}[1]{}
\newcommand{\etal}{{\em et al.}~}

\newcommand{\Exp}{\mathbb{E}}

\DeclareFontFamily{U}{bbold}{}
\DeclareFontShape{U}{bbold}{m}{n}
 {
  <-5.5> s*[1.069] bbold5
  <5.5-6.5> s*[1.069] bbold6
  <6.5-7.5> s*[1.069] bbold7
  <7.5-8.5> s*[1.069] bbold8
  <8.5-9.5> s*[1.069] bbold9
  <9.5-11> s*[1.069] bbold10
  <11-15> s*[1.069] bbold12
  <15-> s*[1.069] bbold17
 }{}

\DeclareRobustCommand{\identity}{%
  \text{\usefont{U}{bbold}{m}{n}1}%
}

\usepackage{hyperref}
\usepackage[capitalise]{cleveref}
\usepackage[english]{babel}

\makeatletter
\def\bbl@set@language#1{%
  \edef\languagename{%
    \ifnum\escapechar=\expandafter`\string#1\@empty
    \else\string#1\@empty\fi}%
  \@ifundefined{babel@language@alias@\languagename}{}{%
    \edef\languagename{\@nameuse{babel@language@alias@\languagename}}%
  }%
  \select@language{\languagename}%
  \expandafter\ifx\csname date\languagename\endcsname\relax\else
    \if@filesw
      \protected@write\@auxout{}{\string\select@language{\languagename}}%
      \bbl@for\bbl@tempa\BabelContentsFiles{%
        \addtocontents{\bbl@tempa}{\xstring\select@language{\languagename}}}%
      \bbl@usehooks{write}{}%
    \fi
  \fi}
\newcommand{\DeclareLanguageAlias}[2]{%
  \global\@namedef{babel@language@alias@#1}{#2}%
}
\makeatother

\DeclareLanguageAlias{en}{english}

\begin{document}
\title{Breaking the Speed Limit for Perfect Quantum State Transfer}
\author{Weichen Xie}\affiliation{Department of Mathematics, Clarkson University, Potsdam, New York, USA 13699-5815.}
\email{Email: xiew@clarkson.edu}
\author{Alastair Kay}
\affiliation{Department of Mathematics, Royal Holloway University of London, Egham, Surrey, TW20 0EX, UK.}
\email{Email: alastair.kay@rhul.ac.uk}
\author{Christino Tamon}
\affiliation{Department of Computer Science, Clarkson University, Potsdam, New York, USA 13699-5815.}
\email{Email: ctamon@clarkson.edu}
\date{\today}
\begin{abstract}
We describe a protocol for perfectly transferring a quantum state from one party to another under the dynamics of a fixed, engineered Hamiltonian. Our protocol combines the concepts of fractional revival, dual rail encoding, and a rare glimpse of the anti-Zeno effect. Remarkably, the transfer happens faster than the speed limit for perfect quantum state transfer \cite{yung2006,kay2016b}. 
\end{abstract}
\maketitle

\section{Introduction}

The current generations of quantum computers are extremely promising. They are at the cusp of proving quantum advantage \cite{arute2019,zhong2020}, although they are not yet capable of implementing arbitrary algorithms of useful sizes. Critical to their performance is the ability to perform as many operations as possible before the inevitable decoherence-induced limits are reached. Theoretically, it is crucial to understand the ultimate bounds -- how long \emph{must} we take to perform a specific operation, and can such a bound be saturated? For example, in the protocol of perfect state transfer, introduced in \cite{bose2003,christandl2004}, there is a speed limit for the protocol: there cannot exist a solution that takes less than a certain time \cite{yung2006,kay2016b}. Hence, the solution in \cite{christandl2004} is optimal, operating at twice the speed of the consecutive swap gates that would be specified in the gate model.

Even more interesting are the conditions under which any no-go theorem is proved and whether, by relaxing any of them, the no-go result can be violated. State transfer, for example, imposes perfect transfer, and requires no time control of the system Hamiltonian. Relaxing either one permits protocols that may still be of practical relevance, but acting faster \cite{apollaro2012,kay2022,murphy2010}. This is crucial for our fledgling quantum computers.

Another potential loophole, that we investigate in this paper, is the assumption of unitarity and, consequently, deterministic arrival. The anti-Zeno effect suggests the possibility that repeated measurement at appropriate times \emph{could} speed up the evolution \cite{harrington2017}. We will demonstrate a protocol, taking inspiration from the heralded, non-deterministic arrival protocols of \cite{burgarth2005b,burgarth2005,burgarth2005a}, improving them for the specific purpose of use with an engineered spin chain. This will allow us to demonstrate a violation of the speed limit thanks to the anti-Zeno effect. The protocol becomes particularly elegant if the engineered chains that we choose exhibit fractional revival \cite{dai2010,kay2010a}, introduced in \cref{sec:FR}.  The main protocol is detailed in \cref{sec:monorail}, where we introduce a monorail encoding, i.e.\ unlike \cite{burgarth2005b,burgarth2005} which required a dual-rail encoding across two parallel spins chains, we only need a single chain. Also, should our state fail to arrive, it is like a complete reset of the protocol, which is computationally far easier than the repeated updates and recalculations required in \cite{burgarth2005b,burgarth2005}. In \cref{sec:genest}, we shall prove that the analytic solutions of \cite{genest2016} saturate an equivalent speed limit for fractional revival, and are hence the optimal choice. We show that, for even length chains, they yield a state transfer protocol whose expected arrival time is less than the fastest possible perfect state transfer time. The advantage is subtle, however, as illustrated by the fact that there is no advantage in the case of odd chain length.

\section{Fractional Revivals}\label{sec:FR}

Let $H$ be the Hamiltonian of an $N$-qubit spin chain,
\[
	H = \frac12\sum_{n=1}^{N-1} J_n(X_n X_{n+1} + Y_n Y_{n+1})-\frac12\sum_{n=1}^{N} B_n Z_n,
\]
where $Z_n$ denotes application of the Pauli-$Z$ matrix on qubit $n$.
The standard concept of perfect state transfer, as introduced in \cite{bose2003, christandl2004,kay2010a}, involves a unitary evolution of an unknown single-qubit state $\ket{\psi}$ from one end of the chain to the other,
\begin{align*}
     e^{-i H \tau}\ket{\psi}\otimes\ket{\underline{0}}= \ket{\underline{0}}\otimes\left(e^{-iZ\phi/2}\ket{\psi}\right)
\end{align*}
in the state transfer time $\tau$, up to a corrective phase rotation by angle $\phi$. Here, we denote $\ket{\underline{0}}=\ket{0}^{\otimes(N-1)}$ and $\ket{\underline{n}}=\ket{0}^{\otimes(n-1)}\ket{1}\ket{0}^{\otimes(N-n)}$. 
The representation of $H$ in the basis $\{\ket{\underline{n}}\}_{n=1}^N$ is denoted by $H_1$, and is an $N\times N$ tridiagonal matrix with ordered eigenvalues $\lambda_n>\lambda_{n+1}$.

It is well-known (see \cite{yung2006,kay2016b}) that on any $N$-qubit spin chain, there is a speed limit to the transfer:
\begin{equation}\label{eq:speed_limit}
	J_{\max}\tau \ge \frac{\pi}{4}\sqrt{N^2 -\frac12(1-(-1)^N)}
\end{equation}
where $J_{\max} = \max_{n} |J_{n}|$ is the maximum coupling strength \footnote{Evaluating the speed limit using $J_{\max}\tau$ removes any ambiguity arising from the fact that for any $H$ that achieves transfer in time $\tau$, $\gamma H$ achieves transfer in time $\tau(\gamma)=\tau/\gamma$.}.

Suppose our spin chain has a {\em fractional revival} (FR).
More specifically, if we can write the evolution as
\begin{align}
e^{-i H \tau_0}\ket{\underline 1} = \cos\theta\ket{\underline1} + \sin\theta e^{i\phi}\ket{\underline N},\label{eq:FR}
\end{align}
up to a global phase,
we say that $H$ has a $\theta$-revival between $1$ and $N$ at time $\tau_0$. The phase $\phi$ is irrelevant to our study (it will typically be $-\frac{\pi}{2}$, and will always be known, so that it can be compensated for). For such a revival, the state does not transfer, but gets entangled between extremal sites:
$$
\ket{\psi}\ket{\underline{0}}\longrightarrow\cos\theta\ket{\psi}\ket{\underline{0}}+\sin\theta\ket{\underline{0}}\left(e^{-iZ\phi/2}\ket{\psi}\right).
$$

Fractional revival chains have some key properties that will be of use to us. In particular, \cref{eq:FR} implies that
\begin{align}\label{eq:all}
e^{-i H \tau_0}\ket{\underline n} = \cos\theta\ket{\underline n} + \sin\theta e^{i\phi}\ket{\underline{N+1-n}}
\end{align}
for any $n\neq \frac{N+1}{2}$. To see this, observe, for example, that $\ket{\underline 2}= (H\ket{\underline 1}-B_1\ket{\underline 1})/J_1$ such that
\begin{align*}
e^{-i H \tau_0}\ket{\underline 2}&=\frac{1}{J_1}(H-B_1\identity)e^{-i H \tau_0}\ket{\underline 1} \\
&=\frac{1}{J_1}(H-B_1\identity)(\cos\theta\ket{\underline 1}+\sin\theta e^{i\phi}\ket{\underline N}) \\
&=\cos\theta\ket{\underline 2}+\sin\theta e^{i\phi}\left(\textstyle\frac{J_{N-1}}{J_1}\ket{\underline{N-1}}+\frac{B_N-B_1}{J_1}\ket{\underline N}\right).
\end{align*}
Since this output must be orthogonal to the output of the evolution \cref{eq:FR}, and must also be normalised, we see that $B_1=B_N$ and $J_1=J_{N-1}$.

Indeed, following through the consequences of this argument shows that if the chain length, $N$, is even, then the chain must be symmetric. If $N$ is odd, it is either a symmetric chain, or the central two coupling strengths $J_{(N\pm 1)/2}$ could be asymmetric. 

In the case of FR on a symmetric chain, we can be more explicit about the phases. A symmetric chain divides into symmetric and anti-symmetric subspaces. These have eigenvalues $\{\lambda_{2n-1}\}$ and $\{\lambda_{2n}\}$ respectively. The initial state $\ket{\underline 1}$ and its $\theta$-revival are both supported only on the state $(\ket{\underline 1}+\ket{\underline N})/\sqrt{2}$ in the symmetric subspace. Hence, within that subspace, we must have a perfect revival of that state (which, in fact, means a perfect revival of all symmetric states), up to a global phase. The same also holds for the antisymmetric space. Thus, the only relevant parameter is the relative phase between these two perfect revivals, which we call $2\theta$. We have
\begin{align*}
\frac{1}{\sqrt{2}}(\ket{\underline 1}+\ket{\underline N})&\mapsto\frac{1}{\sqrt{2}}(\ket{\underline 1}+\ket{\underline N}) \\
\frac{1}{\sqrt{2}}(\ket{\underline 1}-\ket{\underline N})&\mapsto\frac{e^{i2\theta}}{\sqrt{2}}(\ket{\underline 1}-\ket{\underline N}).
\end{align*}
We immediately learn that for a symmetric FR chain,
\begin{equation}\label{eq:fr_evolution}
\ket{\underline{1}}\mapsto\cos\theta\ket{\underline 1}-i\sin\theta\ket{\underline N}, \quad \ket{\underline{N}}\mapsto\cos\theta\ket{\underline N}-i\sin\theta\ket{\underline 1}
\end{equation}
up to a global phase.

\bigskip

\section{All Aboard the Monorail}\label{sec:monorail}

We now come to the key part of our protocol. This has many aspects in common with \cite{burgarth2005b,burgarth2005,burgarth2005a}, which used a dual rail encoding -- an encoding of the qubit entirely within the single-excitation subspace, but split across two parallel spin chains. Here, we will use the same encoding procedure, but on a single rail. Specifically, we take our unknown single-qubit state $\ket{\psi}=\alpha\ket{0}+\beta\ket{1}$, and encode it as $\alpha\ket{\underline 2}+\beta\ket{\underline1}$.

We place this state on the chain, wait the time $\tau_0$, and test to see if the single excitation has arrived without looking at the state itself. As in \cite{burgarth2005b,burgarth2005}, the key to this working is that the arrival amplitude for $1\rightarrow N$ and $2\rightarrow N-1$ should be equal. By construction, \cref{eq:all}, an FR chain at the time $\tau_0$ has this property. If the excitation has arrived, the state has successfully transferred. 
\begin{widetext}
$$
e^{-iH\tau_0}(\alpha\ket{\underline 2}+\beta\ket{\underline 1})=\cos\theta(\alpha\ket{\underline 2}+\beta\ket{\underline 1})+\sin\theta e^{i\phi}(\alpha\ket{\underline{N-1}}+\beta\ket{\underline N})\xrightarrow{\text{\ \ measure\ \ }}\left\{\begin{array}{cc}
\alpha\ket{\underline{N-1}}+\beta\ket{\underline N} & p=\sin^2\theta \\
\alpha\ket{\underline 2}+\beta\ket{\underline 1} &p=\cos^2\theta
\end{array}\right.
$$
\end{widetext}
If the excitation is not present, due to the properties of FR, we know it must be back at the start, and the protocol can just repeat. Repetition until success completes after an expected time
$$
\mathbb{E}(J_{\max}T)=J_{\max}\tau_0\sum_{n=1}^{\infty}n\cos^{2n-2}\theta\sin^2\theta=\frac{J_{\max}\tau_0}{\sin^2\theta}.
$$
Since the success of each repetition is independent, one can apply Chernoff bounds to show that the need for many repetitions vanishes exponentially.

\begin{figure}[!t]
\centering
\begin{quantikz}
\lstick{$\ket{\psi}$} 	& \octrl{1}	& \gate[wires=6][2cm]{FR}	& \qw 	& \qw	& \qw 	& \\
\lstick{$\ket{0}$}	& \targ{}	&				& \qw		& \qw		& \qw 	& \\
\lstick{$\ket{0}$}	& \qw		&				& \qw		& \qw		& \qw 	& \\
\lstick{$\ket{0}$}	& \qw		&				& \qw		& \qw		& \qw 	& \\
\lstick{$\ket{0}$}	& \qw		&				& \targ{}	& \meter{} 	& \cw 	& \rstick{$1$} \\
\lstick{$\ket{0}$} 	& \qw 		&				& \ctrl{-1} & \qw		& \qw 	& \rstick{$\ket{\psi}$}
\end{quantikz}
\caption{PST spin-chain with heralded anti-Zeno effect. If $0$ is measured, we simply wait for the next FR cycle; otherwise
the state $\ket{\psi}$ has arrived.}
\label{fig:zeno-chain}
\end{figure}
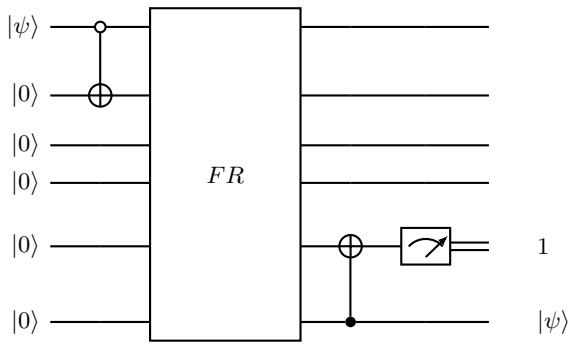

The full protocol is summarised in \cref{fig:zeno-chain}. Including the encoding steps, we have
\begin{enumerate}
	\item Prepare an arbitrary single-qubit state $\ket{\psi}$ in the first qubit of the chain.
\item Apply a controlled-\textsc{not} gate on the first two qubits (with the first qubit being the control, and applying the target operation if the control qubit is in $\ket{0}$).
\item Evolve the FR chain up to its revival time $\tau_0$.
\item Apply the standard c\textsc{not} gate on the last two qubits (with the last qubit being the control).
\item Measure the second last qubit in the $Z$ basis.
\item If the measurement result is $1$, halt (as state $\ket{\psi}$ has arrived on the last qubit).
	Otherwise, repeat from step 3 (as the state is the same as it was then).
\end{enumerate}

\section{Spin chain}\label{sec:genest}

In this section, we show that the Hamiltonian constructed by Genest \etal in Section 6 of \cite{genest2016} for a chain with even $N$ has an expected transfer time that is lower than any deterministic $N$-qubit spin chain with the same maximum coupling strength (edge weight).
The parameters that give a $\theta$-revival at time $\tau_0 = \pi/2$ are, in our indexing convention,
\begin{equation}\label{eq:FRchain}
	J_{n} = \sqrt{\frac{n(N-n)((N-2n)^2 - \frac{4\theta^2}{\pi^2})}{(N-1-2n)(N+1-2n)}},\qquad B_n=0
\end{equation}
We will be particularly interested in
$$
\theta_c=\frac{\pi}{2}\sqrt{1-\frac{3}{N^2-1}},
$$
which is extremely close to possessing perfect state transfer ($\theta=\frac{\pi}{2}$). This choice fixes that $J_{\max}=J_{N/2\pm 1} = J_{N/2} = \theta_c N/\pi$. 

If $T$ is the arrival time of the state (which is a geometric random variable),
the expected transport time is 
\[
	\Exp[J_{\max}T] 
	= \frac{\pi}{2}\frac{\theta N}{\pi}\frac{1}{\sin^2\theta}=\frac{N\theta}{2\sin^2\theta},
\]
which should be compared to the perfect transfer speed limit of \cref{eq:speed_limit}.
At our chosen value of $\theta_c$,
$$
\frac{\Exp[J_{\max}T] }{J_{\max}\tau_{\min}}=\frac{2\theta_c}{\pi\sin^2\theta_c}=\frac{1-\epsilon}{\cos^2\frac{\pi\epsilon}{2}}
$$
where $\theta_c=\frac{\pi}{2}(1-\epsilon)$. This is evidently smaller than 1 for $\epsilon<\frac12$, which is certainly true for $\theta=\theta_c$ and $N>2$.

\subsection{Fastest Possible FR}

We will now justify that the chain we have used, defined by \cref{eq:FRchain}, yields the smallest possible value of $J_{\max}\tau_0$ for any given $\theta$. This proof strongly parallels the speed limit proof of \cite{yung2006}. Since $N$ is even, the chain is symmetric, and the evolution is given by \cref{eq:fr_evolution}. Since the symmetric subspace has perfect revivals in time $\tau_0$, its eigenvalues must satisfy $\lambda_1-\lambda_{2n-1}=\frac{2\pi}{\tau_0}k_n$ for some integer $k_n$. Similarly, the antisymmetric subspace must satisfy $\lambda_1-\lambda_{2n}=\frac{2\pi}{\tau_0}k'_n+\frac{2\theta}{\tau_0}$, in order to get the correct relative phase. Most critically for us,
\begin{equation}\label{eq:gap_bound}
\lambda_{2n-1}-\lambda_{2n}\geq \frac{2\theta}{\tau_0}.
\end{equation}
The matrix $H_1$ has a symmetry $SH_1S=H_1$ where
$$
S=\sum_{n=1}^N\ket{\underline{n}}\bra{\underline{N+1-n}}.
$$
Now,
$$
2J_{\max}\geq 2J_{N/2}=\text{Tr}(H_1S)=\sum_n\lambda_{2n-1}-\lambda_{2n}\geq \frac{N}{2}\frac{2\theta}{\tau_0}
$$
by \cref{eq:gap_bound}. We conclude that
$$
J_{\max}\tau_0\geq \frac{N\theta}{2}.
$$
This is exactly the value that we achieved above. Hence, the solution from \cite{genest2016} is optimal for any given $\theta_c\leq\theta\leq\pi-\theta_c$. This recovers the previously known results for perfect state transfer when $\theta=\pi/2$ \cite{yung2006,christandl2004,kay2010a}.

In addition to the solution for even chain lengths, \cite{genest2016} gives an equivalent solution for odd chain lengths. This works within our protocol (as does any chain with FR), but does not provide any speed advantage over state transfer. Moreover, following \cite{kay2016b}, one can prove that this is also the fastest symmetric chain with a given FR angle between the chain ends. One simply replaces the bounds
\begin{align*}
\lambda_{2n}-\lambda_{2n+1}\geq\frac{\pi}{\tau}&\longrightarrow \lambda_{2n}-\lambda_{2n+1}\geq\frac{2(\pi-\theta)}{\tau_0} \\
\lambda_{2n-1}-\lambda_{2n}\geq\frac{\pi}{\tau}&\longrightarrow \lambda_{2n-1}-\lambda_{2n}\geq\frac{2\theta}{\tau_0}.
\end{align*}
We conclude that
$$
J_{\max}\tau_0\geq\frac{\sqrt{(N^2-1)\theta(\pi-\theta)}}{2},
$$
which is saturated by the solution in \cite{genest2016} for any $\theta$ such that the largest coupling is the central one.

When $N$ is odd, solutions are not required to be symmetric. In the asymmetric case, a chain with a $\theta$-revival is symmetric apart from the middle two coupling strengths 
$J_{(N-1)/2}=\sqrt{2}J\cos\eta$ and $J_{(N+1)/2}=\sqrt{2}J\sin\eta$. $H_1$ is similar to
$
H'=UH_1U^\dagger
$
with
\begin{multline*}
U=\proj{\underline{\frac{N+1}{2}}}+\\\sum_{n=1}^{(N-1)/2}\sin\left(\frac{\pi}{4}-\eta\right)(\proj{\underline{n}}-\proj{\underline{N+1-n}})\\ +\cos\left(\frac{\pi}{4}-\eta\right)(\ket{\underline{N+1-n}}\bra{\underline{n}}+\ket{\underline{n}}\bra{\underline{N+1-n}}).
\end{multline*}
$H'$ is a symmetric chain with central coupling $J$, and all other parameters equal to those of $H_1$. The evolution of the two are related by 
$$
e^{-iH_1\tau_0}\ket{\underline 1}=U^\dagger e^{-iH'\tau_0}U\ket{\underline{1}}.
$$
These can only be equal if $H'$ possesses FR by some angle $\theta'$. Thus, the right-hand side is entirely known thanks to \cref{eq:fr_evolution}. Hence, the two FR angles are related by
$$
\sin\theta=\sin(2\eta)\sin\theta'.
$$
The expected state transfer time can then be evaluated relative to that of the symmetric chain,
$$
\mathbb{E}[J_{\max}T] = \mathbb{E}[J_{\max}'T']\times\left\{\begin{array}{cc}
\frac{\sqrt{2}\cos\eta}{\sin^22\eta}& J_{\max}=\sqrt{2}J\cos\eta \\
\frac{\sqrt{2}\sin\eta}{\sin^22\eta}& J_{\max}=\sqrt{2}J\sin\eta \\
\frac{1}{\sin^22\eta} & \text{otherwise}
\end{array}\right.
%\frac{\sqrt{2}\cos\eta J_{\max}\tau_0}{\sin^2\theta'} = \frac{\sqrt{2}\cos\eta}{\sin^2 2\eta} \mathbb{E}[J_{\max}T], }
$$
In all these cases, $\mathbb{E}[J_{\max}T]>\mathbb{E}[J_{\max}'T']$ -- the symmetric solution has the best expected transfer time, which we have already shown cannot break the speed limit.

\section{Conclusions}

In this work, we have constructed a probabilistic protocol for perfect state transfer
with provable speedup over {\em any} deterministic perfect state transfer chain with
the same maximum coupling strength.

The speedup that we achieve is miniscule, only reducing the state transfer time from $\tau$ by an amount $O(\tau/N^2)$, at the cost of non-deterministic arrival. If we account fully for the time required for the decoding operation before measurement, this outweighs the speedup! Nevertheless, our point is that the speed limit \emph{can} be broken thanks to the anti-Zeno effect. The FR provides a particularly elegant method allowing us to compress the dual-rail encoding onto a single chain, and also yielding information about what state the chain is reset to in the case of failure. Its disadvantage is that, after a failure, the state is reset such that it has to travel the full length of the chain again before the next possible arrival, which is why the use of FR for values of $\theta$ far from perfect state transfer is impractical. There may be other systems which are more efficient from this perspective, and yield greater advantages.

There are several useful properties that are built in to the protocol. While we have dealt with the specific case that the rest of the chain is initialised as $\ket{\underline 0}$, our encoding is exactly that used to tolerate any initial state \cite{kay2010a,albanese2004}, and hence this carries over immediately so long as the sites $N$ and $N-1$ are prepared in $\ket{0}$. Equally, if we have more information to transport, such as a qutrit, one simply uses a superposition over the states $\{\ket{\underline 1},\ket{\underline 2},\ket{\underline 3}\}$, and detects arrival by looking for the presence of a $\ket{1}$ on the final three sites. This requires a minimal update to our encoding/decoding procedures.

\section*{Acknowledgments}

Work done while the authors were attending the workshop 
``Graph Theory, Algebraic Combinatorics, and Mathematical Physics'' 
at Centre de Recherches Math\'{e}matiques (CRM), Universit\'{e} de Montr\'{e}al.
C.T.\ would like to thank CRM for its hospitality and support during his sabbatical visit.
W.X.\ was supported by NSF travel grant DMS-2212755.
We thank Luc Vinet and Ada Chan for helpful discussions.

%\bibliography{References}

\begin{thebibliography}{18}%
\makeatletter
\providecommand \@ifxundefined [1]{%
 \@ifx{#1\undefined}
}%
\providecommand \@ifnum [1]{%
 \ifnum #1\expandafter \@firstoftwo
 \else \expandafter \@secondoftwo
 \fi
}%
\providecommand \@ifx [1]{%
 \ifx #1\expandafter \@firstoftwo
 \else \expandafter \@secondoftwo
 \fi
}%
\providecommand \natexlab [1]{#1}%
\providecommand \enquote  [1]{``#1''}%
\providecommand \bibnamefont  [1]{#1}%
\providecommand \bibfnamefont [1]{#1}%
\providecommand \citenamefont [1]{#1}%
\providecommand \href@noop [0]{\@secondoftwo}%
\providecommand \href [0]{\begingroup \@sanitize@url \@href}%
\providecommand \@href[1]{\@@startlink{#1}\@@href}%
\providecommand \@@href[1]{\endgroup#1\@@endlink}%
\providecommand \@sanitize@url [0]{\catcode `\\12\catcode `\$12\catcode
  `\&12\catcode `\#12\catcode `\^12\catcode `\_12\catcode `\%12\relax}%
\providecommand \@@startlink[1]{}%
\providecommand \@@endlink[0]{}%
\providecommand \url  [0]{\begingroup\@sanitize@url \@url }%
\providecommand \@url [1]{\endgroup\@href {#1}{\urlprefix }}%
\providecommand \urlprefix  [0]{URL }%
\providecommand \Eprint [0]{\href }%
\providecommand \doibase [0]{https://doi.org/}%
\providecommand \selectlanguage [0]{\@gobble}%
\providecommand \bibinfo  [0]{\@secondoftwo}%
\providecommand \bibfield  [0]{\@secondoftwo}%
\providecommand \translation [1]{[#1]}%
\providecommand \BibitemOpen [0]{}%
\providecommand \bibitemStop [0]{}%
\providecommand \bibitemNoStop [0]{.\EOS\space}%
\providecommand \EOS [0]{\spacefactor3000\relax}%
\providecommand \BibitemShut  [1]{\csname bibitem#1\endcsname}%
\let\auto@bib@innerbib\@empty
%</preamble>
\bibitem [{\citenamefont {Yung}(2006)}]{yung2006}%
  \BibitemOpen
  \bibfield  {author} {\bibinfo {author} {\bibfnamefont {M.-H.}\ \bibnamefont
  {Yung}},\ }\bibfield  {title} {\bibinfo {title} {Quantum speed limit for
  perfect state transfer in one dimension},\ }\href
  {https://doi.org/10.1103/PhysRevA.74.030303} {\bibfield  {journal} {\bibinfo
  {journal} {Phys. Rev. A}\ }\textbf {\bibinfo {volume} {74}},\ \bibinfo
  {pages} {030303(R)} (\bibinfo {year} {2006})}\BibitemShut {NoStop}%
\bibitem [{\citenamefont {Kay}(2016)}]{kay2016b}%
  \BibitemOpen
  \bibfield  {author} {\bibinfo {author} {\bibfnamefont {A.}~\bibnamefont
  {Kay}},\ }\bibfield  {title} {\bibinfo {title} {A {{Note}} on the {{Speed}}
  of {{Perfect State Transfer}}},\ }\href@noop {} {\bibfield  {journal}
  {\bibinfo  {journal} {arXiv}\ } (\bibinfo {year} {2016})},\ \Eprint
  {https://arxiv.org/abs/1609.01854} {arXiv:1609.01854} \BibitemShut {NoStop}%
\bibitem [{\citenamefont {Arute}\ and\ \citenamefont
  {{others}}(2019)}]{arute2019}%
  \BibitemOpen
  \bibfield  {author} {\bibinfo {author} {\bibfnamefont {F.}~\bibnamefont
  {Arute}}\ and\ \bibinfo {author} {\bibnamefont {{others}}},\ }\bibfield
  {title} {\bibinfo {title} {Quantum supremacy using a programmable
  superconducting processor},\ }\href
  {https://doi.org/10.1038/s41586-019-1666-5} {\bibfield  {journal} {\bibinfo
  {journal} {Nature}\ }\textbf {\bibinfo {volume} {574}},\ \bibinfo {pages}
  {505} (\bibinfo {year} {2019})}\BibitemShut {NoStop}%
\bibitem [{\citenamefont {Zhong}\ and\ \citenamefont
  {{others}}(2020)}]{zhong2020}%
  \BibitemOpen
  \bibfield  {author} {\bibinfo {author} {\bibfnamefont {H.-S.}\ \bibnamefont
  {Zhong}}\ and\ \bibinfo {author} {\bibnamefont {{others}}},\ }\bibfield
  {title} {\bibinfo {title} {Quantum computational advantage using photons},\
  }\href {https://doi.org/10.1126/science.abe8770} {\bibfield  {journal}
  {\bibinfo  {journal} {Science}\ }\textbf {\bibinfo {volume} {370}},\ \bibinfo
  {pages} {1460} (\bibinfo {year} {2020})}\BibitemShut {NoStop}%
\bibitem [{\citenamefont {Bose}(2003)}]{bose2003}%
  \BibitemOpen
  \bibfield  {author} {\bibinfo {author} {\bibfnamefont {S.}~\bibnamefont
  {Bose}},\ }\bibfield  {title} {\bibinfo {title} {Quantum {{Communication}}
  through an {{Unmodulated Spin Chain}}},\ }\href
  {https://doi.org/10.1103/PhysRevLett.91.207901} {\bibfield  {journal}
  {\bibinfo  {journal} {Phys. Rev. Lett.}\ }\textbf {\bibinfo {volume} {91}},\
  \bibinfo {pages} {207901} (\bibinfo {year} {2003})}\BibitemShut {NoStop}%
\bibitem [{\citenamefont {Christandl}\ \emph {et~al.}(2004)\citenamefont
  {Christandl}, \citenamefont {Datta}, \citenamefont {Ekert},\ and\
  \citenamefont {Landahl}}]{christandl2004}%
  \BibitemOpen
  \bibfield  {author} {\bibinfo {author} {\bibfnamefont {M.}~\bibnamefont
  {Christandl}}, \bibinfo {author} {\bibfnamefont {N.}~\bibnamefont {Datta}},
  \bibinfo {author} {\bibfnamefont {A.}~\bibnamefont {Ekert}},\ and\ \bibinfo
  {author} {\bibfnamefont {A.~J.}\ \bibnamefont {Landahl}},\ }\bibfield
  {title} {\bibinfo {title} {Perfect {{State Transfer}} in {{Quantum Spin
  Networks}}},\ }\href {https://doi.org/10.1103/PhysRevLett.92.187902}
  {\bibfield  {journal} {\bibinfo  {journal} {Phys. Rev. Lett.}\ }\textbf
  {\bibinfo {volume} {92}},\ \bibinfo {pages} {187902} (\bibinfo {year}
  {2004})}\BibitemShut {NoStop}%
\bibitem [{\citenamefont {Apollaro}\ \emph {et~al.}(2012)\citenamefont
  {Apollaro}, \citenamefont {Banchi}, \citenamefont {Cuccoli}, \citenamefont
  {Vaia},\ and\ \citenamefont {Verrucchi}}]{apollaro2012}%
  \BibitemOpen
  \bibfield  {author} {\bibinfo {author} {\bibfnamefont {T.~J.~G.}\
  \bibnamefont {Apollaro}}, \bibinfo {author} {\bibfnamefont {L.}~\bibnamefont
  {Banchi}}, \bibinfo {author} {\bibfnamefont {A.}~\bibnamefont {Cuccoli}},
  \bibinfo {author} {\bibfnamefont {R.}~\bibnamefont {Vaia}},\ and\ \bibinfo
  {author} {\bibfnamefont {P.}~\bibnamefont {Verrucchi}},\ }\bibfield  {title}
  {\bibinfo {title} {99\%-fidelity ballistic quantum-state transfer through
  long uniform channels},\ }\href {https://doi.org/10.1103/PhysRevA.85.052319}
  {\bibfield  {journal} {\bibinfo  {journal} {Phys. Rev. A}\ }\textbf {\bibinfo
  {volume} {85}},\ \bibinfo {pages} {052319} (\bibinfo {year}
  {2012})}\BibitemShut {NoStop}%
\bibitem [{\citenamefont {Kay}(2022)}]{kay2022}%
  \BibitemOpen
  \bibfield  {author} {\bibinfo {author} {\bibfnamefont {A.}~\bibnamefont
  {Kay}},\ }\href {https://doi.org/10.48550/arXiv.2207.01954} {\bibinfo {title}
  {Incorporating {{Encoding}} into {{Quantum System Design}}}} (\bibinfo {year}
  {2022}),\ \Eprint {https://arxiv.org/abs/2207.01954} {arXiv:2207.01954
  [quant-ph]} \BibitemShut {NoStop}%
\bibitem [{\citenamefont {Murphy}\ \emph {et~al.}(2010)\citenamefont {Murphy},
  \citenamefont {Montangero}, \citenamefont {Giovannetti},\ and\ \citenamefont
  {Calarco}}]{murphy2010}%
  \BibitemOpen
  \bibfield  {author} {\bibinfo {author} {\bibfnamefont {M.}~\bibnamefont
  {Murphy}}, \bibinfo {author} {\bibfnamefont {S.}~\bibnamefont {Montangero}},
  \bibinfo {author} {\bibfnamefont {V.}~\bibnamefont {Giovannetti}},\ and\
  \bibinfo {author} {\bibfnamefont {T.}~\bibnamefont {Calarco}},\ }\bibfield
  {title} {\bibinfo {title} {Communication at the quantum speed limit along a
  spin chain},\ }\href {https://doi.org/10.1103/PhysRevA.82.022318} {\bibfield
  {journal} {\bibinfo  {journal} {Phys. Rev. A}\ }\textbf {\bibinfo {volume}
  {82}},\ \bibinfo {pages} {022318} (\bibinfo {year} {2010})}\BibitemShut
  {NoStop}%
\bibitem [{\citenamefont {Harrington}\ \emph {et~al.}(2017)\citenamefont
  {Harrington}, \citenamefont {Monroe},\ and\ \citenamefont
  {Murch}}]{harrington2017}%
  \BibitemOpen
  \bibfield  {author} {\bibinfo {author} {\bibfnamefont {P.~M.}\ \bibnamefont
  {Harrington}}, \bibinfo {author} {\bibfnamefont {J.~T.}\ \bibnamefont
  {Monroe}},\ and\ \bibinfo {author} {\bibfnamefont {K.~W.}\ \bibnamefont
  {Murch}},\ }\bibfield  {title} {\bibinfo {title} {Quantum {{Zeno Effects}}
  from {{Measurement Controlled Qubit-Bath Interactions}}},\ }\href
  {https://doi.org/10.1103/PhysRevLett.118.240401} {\bibfield  {journal}
  {\bibinfo  {journal} {Phys. Rev. Lett.}\ }\textbf {\bibinfo {volume} {118}},\
  \bibinfo {pages} {240401} (\bibinfo {year} {2017})}\BibitemShut {NoStop}%
\bibitem [{\citenamefont {Burgarth}\ and\ \citenamefont
  {Bose}(2005{\natexlab{a}})}]{burgarth2005b}%
  \BibitemOpen
  \bibfield  {author} {\bibinfo {author} {\bibfnamefont {D.}~\bibnamefont
  {Burgarth}}\ and\ \bibinfo {author} {\bibfnamefont {S.}~\bibnamefont
  {Bose}},\ }\bibfield  {title} {\bibinfo {title} {Conclusive and arbitrarily
  perfect quantum state transfer using parallel spin chain channels},\ }\href
  {https://doi.org/10.1103/PhysRevA.71.052315} {\bibfield  {journal} {\bibinfo
  {journal} {Phys. Rev. A}\ }\textbf {\bibinfo {volume} {71}},\ \bibinfo
  {pages} {052315} (\bibinfo {year} {2005}{\natexlab{a}})},\ \Eprint
  {https://arxiv.org/abs/quant-ph/0406112} {arXiv:quant-ph/0406112}
  \BibitemShut {NoStop}%
\bibitem [{\citenamefont {Burgarth}\ and\ \citenamefont
  {Bose}(2005{\natexlab{b}})}]{burgarth2005}%
  \BibitemOpen
  \bibfield  {author} {\bibinfo {author} {\bibfnamefont {D.}~\bibnamefont
  {Burgarth}}\ and\ \bibinfo {author} {\bibfnamefont {S.}~\bibnamefont
  {Bose}},\ }\bibfield  {title} {\bibinfo {title} {Perfect quantum state
  transfer with randomly coupled quantum chains},\ }\href
  {https://doi.org/10.1088/1367-2630/7/1/135} {\bibfield  {journal} {\bibinfo
  {journal} {New J. Phys.}\ }\textbf {\bibinfo {volume} {7}},\ \bibinfo {pages}
  {135} (\bibinfo {year} {2005}{\natexlab{b}})}\BibitemShut {NoStop}%
\bibitem [{\citenamefont {Burgarth}\ \emph {et~al.}(2005)\citenamefont
  {Burgarth}, \citenamefont {Giovannetti},\ and\ \citenamefont
  {Bose}}]{burgarth2005a}%
  \BibitemOpen
  \bibfield  {author} {\bibinfo {author} {\bibfnamefont {D.}~\bibnamefont
  {Burgarth}}, \bibinfo {author} {\bibfnamefont {V.}~\bibnamefont
  {Giovannetti}},\ and\ \bibinfo {author} {\bibfnamefont {S.}~\bibnamefont
  {Bose}},\ }\bibfield  {title} {\bibinfo {title} {Efficient and perfect state
  transfer in quantum chains},\ }\href
  {https://doi.org/10.1088/0305-4470/38/30/013} {\bibfield  {journal} {\bibinfo
   {journal} {J. Phys. A: Math. Gen.}\ }\textbf {\bibinfo {volume} {38}},\
  \bibinfo {pages} {6793} (\bibinfo {year} {2005})}\BibitemShut {NoStop}%
\bibitem [{\citenamefont {Dai}\ \emph {et~al.}(2010)\citenamefont {Dai},
  \citenamefont {Feng},\ and\ \citenamefont {Kwek}}]{dai2010}%
  \BibitemOpen
  \bibfield  {author} {\bibinfo {author} {\bibfnamefont {L.}~\bibnamefont
  {Dai}}, \bibinfo {author} {\bibfnamefont {Y.~P.}\ \bibnamefont {Feng}},\ and\
  \bibinfo {author} {\bibfnamefont {L.~C.}\ \bibnamefont {Kwek}},\ }\bibfield
  {title} {\bibinfo {title} {Engineering quantum cloning through maximal
  entanglement between boundary qubits in an open spin chain},\ }\href
  {https://doi.org/10.1088/1751-8113/43/3/035302} {\bibfield  {journal}
  {\bibinfo  {journal} {J. Phys. A: Math. Theor.}\ }\textbf {\bibinfo {volume}
  {43}},\ \bibinfo {pages} {035302} (\bibinfo {year} {2010})}\BibitemShut
  {NoStop}%
\bibitem [{\citenamefont {Kay}(2010)}]{kay2010a}%
  \BibitemOpen
  \bibfield  {author} {\bibinfo {author} {\bibfnamefont {A.}~\bibnamefont
  {Kay}},\ }\bibfield  {title} {\bibinfo {title} {A {{Review}} of {{Perfect
  State Transfer}} and its {{Application}} as a {{Constructive Tool}}},\ }\href
  {https://doi.org/10.1142/S0219749910006514} {\bibfield  {journal} {\bibinfo
  {journal} {Int. J. Quantum Inform.}\ }\textbf {\bibinfo {volume} {8}},\
  \bibinfo {pages} {641} (\bibinfo {year} {2010})}\BibitemShut {NoStop}%
\bibitem [{\citenamefont {Genest}\ \emph {et~al.}(2016)\citenamefont {Genest},
  \citenamefont {Vinet},\ and\ \citenamefont {Zhedanov}}]{genest2016}%
  \BibitemOpen
  \bibfield  {author} {\bibinfo {author} {\bibfnamefont {V.~X.}\ \bibnamefont
  {Genest}}, \bibinfo {author} {\bibfnamefont {L.}~\bibnamefont {Vinet}},\ and\
  \bibinfo {author} {\bibfnamefont {A.}~\bibnamefont {Zhedanov}},\ }\bibfield
  {title} {\bibinfo {title} {Quantum spin chains with fractional revival},\
  }\href {https://doi.org/10.1016/j.aop.2016.05.009} {\bibfield  {journal}
  {\bibinfo  {journal} {Annals of Physics}\ }\textbf {\bibinfo {volume}
  {371}},\ \bibinfo {pages} {348} (\bibinfo {year} {2016})}\BibitemShut
  {NoStop}%
\bibitem [{Note1()}]{Note1}%
  \BibitemOpen
  \bibinfo {note} {Evaluating the speed limit using $J_{\protect \qopname
  \relax m{max}}\tau $ removes any ambiguity arising from the fact that for any
  $H$ that achieves transfer in time $\tau $, $\gamma H$ achieves transfer in
  time $\tau (\gamma )=\tau /\gamma $.}\BibitemShut {Stop}%
\bibitem [{\citenamefont {Albanese}\ \emph {et~al.}(2004)\citenamefont
  {Albanese}, \citenamefont {Christandl}, \citenamefont {Datta},\ and\
  \citenamefont {Ekert}}]{albanese2004}%
  \BibitemOpen
  \bibfield  {author} {\bibinfo {author} {\bibfnamefont {C.}~\bibnamefont
  {Albanese}}, \bibinfo {author} {\bibfnamefont {M.}~\bibnamefont
  {Christandl}}, \bibinfo {author} {\bibfnamefont {N.}~\bibnamefont {Datta}},\
  and\ \bibinfo {author} {\bibfnamefont {A.}~\bibnamefont {Ekert}},\ }\bibfield
   {title} {\bibinfo {title} {Mirror {{Inversion}} of {{Quantum States}} in
  {{Linear Registers}}},\ }\href
  {https://doi.org/10.1103/PhysRevLett.93.230502} {\bibfield  {journal}
  {\bibinfo  {journal} {Phys. Rev. Lett.}\ }\textbf {\bibinfo {volume} {93}},\
  \bibinfo {pages} {230502} (\bibinfo {year} {2004})}\BibitemShut {NoStop}%
\end{thebibliography}


\end{document}